\documentclass[review]{elsarticle}

\usepackage{lineno,hyperref}
\usepackage[margin = 1.5in]{geometry}

\usepackage{amsmath,amssymb,amsfonts}
\usepackage{algorithmic}
\usepackage{graphicx}
\usepackage{textcomp}
\usepackage{xcolor}
\usepackage{makecell}
\usepackage{amsmath}
\usepackage{multicol, blindtext}
\usepackage{multirow}
\usepackage{tabularx}
\usepackage{csquotes}
\usepackage{float}
\usepackage{bm}
\usepackage{amssymb}
\usepackage{diagbox}
\newcolumntype{C}[1]{>{\centering\arraybackslash}p{#1}}

\graphicspath{ {./Figures/} }
\def\BibTeX{{\rm B\kern-.05em{\sc i\kern-.025em b}\kern-.08em
    T\kern-.1667em\lower.7ex\hbox{E}\kern-.125emX}}

\usepackage{caption}

\journal{Journal of Power Sources}

\begin{document}

\begin{frontmatter}

\title{Online State Estimation for a Physics-Based Lithium-Sulfur Battery Model}

\author{Chu Xu, Timothy Cleary, Daiwei Wang, Guoxing Li, Christopher Rahn, Donghai Wang, Rajesh Rajamani and Hosam K. Fathy \fnref{myfootnote}}
\fntext[myfootnote]{Chu Xu (chuxu88@umd.edu) is a graduate student and Hosam K. Fathy (hfathy@umd.edu, the corresponding author) is a professor of Mechanical Engineering at the University of Maryland, College Park, MD, 27042. Timothy Cleary (tdc142@psu.edu) and Daiwei Wang (dkw5208@psu.edu) are graduate students of Mechanical Engineering at Penn State University. Guoxing Li (gxli@sdu.edu.cn) is a professor at The Science Center for Material Creation and Energy Conversion, Institute of Frontier and Interdisciplinary Science, Shandong University, Qingdao, China. Donghai Wang (dwang@psu.edu) and Christopher Rahn (cdr10@psu.edu) are professors of Mechanical Engineering at Penn State University. Rajesh Rajamani (rajamani@umn.edu) is a professor of Mechanical Engineering at the University of Minnesota.}




\begin{abstract}
This article examines the problem of Lithium-Sulfur (Li-S) battery state estimation. Such estimation is important for the online management of this energy-dense chemistry. The literature uses equivalent circuit models (ECMs) for Li-S state estimation. This article’s main goal is to perform estimation using a physics-based model instead. This approach is attractive because it furnishes online estimates of the masses of individual species in a given Li-S cell. The estimation is performed using an experimentally-validated, computationally tractable zero-dimensional model. Reformulation converts this model from differential algebraic equations (DAEs) to ordinary differential equations (ODEs), simplifying the estimation problem. The article’s first contribution is to show that this model has poor observability, especially in the low plateau region, where the low sensitivity of cell voltage to precipitated sulfur mass complicates the estimation of this mass. The second contribution is to exploit mass conservation to derive a reduced-order model with attractive observability properties in both high and low plateau regions. The final contribution is to use an unscented Kalman filter (UKF) for estimating internal Li-S battery states, while taking constraints on species masses into account. Consistent with the article’s observability analysis, the UKF achieves better low-plateau estimation accuracy when the reduced-order model is used. 
\end{abstract}

\begin{keyword}
Li-S battery\sep observability\sep state estimation\sep unscented Kalman filter
\end{keyword}

\end{frontmatter}



\section{Introduction}
This article examines the problem of using input/output measurements of current and voltage to estimate the internal state of a Lithium-Sulfur (Li-S) battery. State estimation is a broad term from the control systems literature that refers to the estimation of unknown time-varying quantities governing the behavior of a dynamic system. Examples of state estimation problems include estimating a given battery's state of charge (SOC), state of health (SOH), and state of power (SOP).  The main goal of this article is to estimate the masses of the different species that participate in Li-S battery redox reactions. Species mass estimation is valuable because it provides a more detailed picture of what is occurring inside an electrochemical battery compared to, say, SOC estimation alone. 

The motivation for Li-S state estimation is twofold. First, the Li-S chemistry is \textit{important} because it provides a theoretical specific capacity of 1672  $Ah/kg$, significantly higher than more traditional Lithium-ion chemistries~\cite{bresser2013recent,manthiram2014rechargeable, zhang2017advances}. This makes Li-S batteries potentially attractive for applications requiring high specific energies. Second, the Li-S chemistry exhibits \textit{fundamentally different} behaviors compared to more traditional Lithium-ion batteries, including self-discharge through the shuttle effect. Fundamental insights from the existing Lithium-ion battery state estimation literature are, therefore, not always directly applicable to the Li-S state estimation problem. Consider, for example, the question of whether or not the internal state of an electrochemical battery is observable (i.e., can be estimated from input-output data). The answer to this question depends partly on the given battery's internal dynamics, and can therefore change significantly based on whether or not the battery experiences self-discharge. 

There is an extensive literature on Li-S batteries, with at least two broad focus areas. First, the literature examines the problem of optimizing the underlying materials and chemistries in Li-S batteries. Much of this literature focuses on addressing key challenges such as improving cycle life and inhibiting self-discharge~\cite{wang2011graphene,evers2013new,song2013lithium,song2015strong,yuan2016powering}, based on a fundamental understanding of the underlying reactions in Li-S batteries~\cite{kumaresan2008mathematical, fronczek2013insight, hofmann2014mechanistic, ren2016modeling, andrei2018theoretical,danner2019influence,marinescu2016zero,zhang2015modeling}. Second, the literature also examines the problem of designing Li-S battery management systems (BMSs). An effective BMS is important for protecting cells from damage, prolonging battery cycle life, and increasing battery performance metrics such as output power. Typical components of a BMS include a computationally affordable model, a state estimator, and ultimately an optimal control strategy. This article focuses on state estimation, a key element within battery management system design. 

Estimating the state of a dynamic system requires a representation, or model, of the system's underlying dynamics. This representation or model can be derived from the fundamental laws of electrochemistry. It can also be fitted to experimental data using either equivalent circuit methods or machine learning. The literature provides at least three different types of Li-S battery models, as discussed in \cite{fotouhi2017lithium_review}, namely: equivalent circuit models (ECMs)~\cite{kolosnitsyn2011study, deng2013electrochemical, knap2015electrical,propp2016multi}, zero-dimensional electrochemical models~\cite{marinescu2016zero,zhang2015modeling}, and spatially-distributed electrochemical models~\cite{kumaresan2008mathematical, fronczek2013insight, hofmann2014mechanistic, ren2016modeling, andrei2018theoretical,danner2019influence}. These models fall on a spectrum of fidelity and complexity levels. Spatially-distributed electrochemical models, for instance, have the advantage of providing higher-fidelity representations of the underlying battery physics compared to equivalent circuit models, at the cost of higher computational complexity. Spatially-invariant zero-dimensional models provide an attractive middle ground between these two extremes by modeling the underlying redox reactions in Li-S batteries while minimizing computational complexity. 

Existing research on Li-S battery state estimation relies predominantly on either equivalent circuit models or machine learning methods or both. For example, state estimation techniques have been applied to Li-S ECMs in \cite{propp2017kalman, fotouhi2017lithium_sul,propp2019improved} and to machine learning models in~\cite{wang2020state}. In \cite{propp2017kalman}, the extended Kalman filtering (EKF), unscented Kalman filtering (UKF) and particle filtering techniques are applied and compared for experimental Li-S SOC estimation. In \cite{fotouhi2017lithium_sul}, an adaptive neuro-fuzzy inference systems algorithm is developed to estimate the SOC based on real-time cell model ECM parameterization. In \cite{propp2019improved}, a dual Kalman filtering technique is used for combined Li-S state and parameter estimation, Finally, in \cite{wang2020state}, a Long Short-Term Memory Recurrent Neural Network model is built and calibrated for online Li-S state estimation. 

The above literature, while encouraging, does not address the problem of estimating the masses of the various species participating in Li-S battery redox reactions. This is an important problem because unlike traditional Lithium-ion batteries, where Lithium intercalates into and out of the cathode/anode materials, the Li-S chemistry involves multiple reduction reactions that convert $S_8$ gradually to $S^{2-}$ during discharge. This makes the definition of SOC in Li-S batteries a little ambiguous, in the sense that one can potentially define multiple ``states of charge" associated with different reacting species. One possible solution to this problem is: instead of estimating a single overall SOC, one can estimate internal state variables such as the active masses of dissolved sulfur species using a physics-based model. Such state estimation provides a more detailed picture of the internal state of the battery. This, in turn, is potentially useful for predicting and managing phenomena such as the dependence of Li-S discharge capacity on applied current~\cite{wild2019lithium}. 

The main goal of this article is to use a zero-dimensional, physics-based model for the online estimation of species masses in an Li-S battery. The particular model used in this paper builds on earlier research in the literature, including previous work by the authors on the experimental parameterization and validation of the model. The model consists of both differential equations and algebraic constraints, but a reformulation casts it into the more commonly-used state-space form, thereby simplifying the estimation problem.  To the best of the authors' knowledge, this article's use of a zero-dimensional, physics-based model for online species mass estimation is a novel contribution to the literature. Key elements of this contribution include: (i) an observability analysis for the selected Li-S battery model, (ii) the use of model reduction to improve this model's observabiltiy, especially in the low plateau region, and (iii) the development of the desired Li-S state estimator using unscented Kalman filtering. 

The remainder of this article is organized as follows. Section 2 describes the zero dimensional model structure adopted from \cite{marinescu2016zero, zhang2015modeling}, and reformulates this model from a set of differential algebraic equations (DAEs) to a set of ordinary differential equations (ODEs). Section 3 analyzes the model's observability using the empirical observability gramian, and relates the observability gramian to Fisher information matrix in order to obtain the best achievable estimation error bounds on initial states. Moreover, when mass conservation is considered, the number of state variables can be reduced by two and the observability is improved, especially in the low plateau region. Section 4 presents and discusses the simulation results of the UKF-based state estimation for both the full-order and reduced-order ODE models. Finally, Section 5 summarizes the article's conclusions. 

\section{Lithium-Sulfur Battery Model}

A typical Li-S battery discharge voltage curve consists of high and low plateaus. In the high plateau region, the active material, $S_8$, in the cathode accepts electrons to produce the polysulfide, $S_x^{2-}$ ($x$ can be 8, 6, 4). Further polysulfide reduction takes place in the lower voltage plateau region \cite{wild2015lithium}. In parallel, lithium is oxidized in the anode to furnish lithium ions. This section presents a zero-dimensional Li-S battery model, based on earlier research in the literature~\cite{marinescu2016zero,zhang2015modeling}. The model makes the following assumptions: (i) there is an unlimited lithium supply in the cell with a negligible overpotential on the anode side, as in \cite{fronczek2013insight}; (ii) the shuttle effect of polysulfides is not included due to this article's focus on the voltage performance instead of capacity fade \cite{zhang2015modeling}; (iii) only the lowest polysulfide's precipitation reaction ($2Li^{+}+S^{2-} \rightleftharpoons Li_2S \downarrow$) is modeled \cite{zhang2015modeling,marinescu2016zero}; and (iv) only the redox reactions in Table \ref{tab:Reactions} are considered. Although there exist different simplifications of the reaction chain on the cathode side, such as the two-step reduction $S_8 \rightarrow S_4^{2-} \rightarrow S^{2-}$ \cite{danner2015modeling} and the four-step reduction $S_8 \rightarrow S_8^{2-} \rightarrow S_6^{2-}\rightarrow S_4^{2-} \rightarrow S^{2-}$ \cite{ren2016modeling}, our choice of the reaction chain is based on previous work by the authors~\cite{Xu2020parameterid}. This previous work involved both estimating this model's parameters from experimental cycling data and validating the model's accuracy in capturing Li-S battery discharge dynamics.

\begin{table}
\captionsetup{font=scriptsize}
	\begin{center}
		\caption{Reactions and Dissolved Species Considered in the Model}
		\label{tab:Reactions}
		\begin{tabular}{|C{3cm}|C{4cm}|}
			\hline 
			\textbf{Indices} & \textbf{Reactions and Species} \\
			\hline 
			\multirow{4}{*}{ \makecell{Reaction Index\\ j = 1, 2, 3, 4}}  
			& $ \frac{1}{2} S_8 \text{~~} +e^- \rightleftharpoons \frac{1}{2}  S_8^{2-}  $ \\
			& $ \frac{3}{2} S_8^{2-}  +e^- \rightleftharpoons 2 S_6^{2-}  $  \\
			& $  \text{~~} S_6^{2-}  +e^- \rightleftharpoons \frac{3}{2} S_4^{2-}  $  \\
			& $ \frac{1}{6} S_4^{2-} +e^- \rightleftharpoons \frac{2}{3}  S^{2-}  $  \\
			\hline
		 \makecell{Reaction Index \\i = 1, 2, ..., 5} & $S_8,~S_8^{2-}, S_6^{2-},S_4^{2-},S^{2-}$  \\	
		\hline
		\end{tabular}
	\end{center}
\end{table}

\subsection{Model Derivation}
\begin{figure*}
    \centering
    \includegraphics[trim={0.5cm 4.4cm 2cm 2cm},clip,width= 1\textwidth]{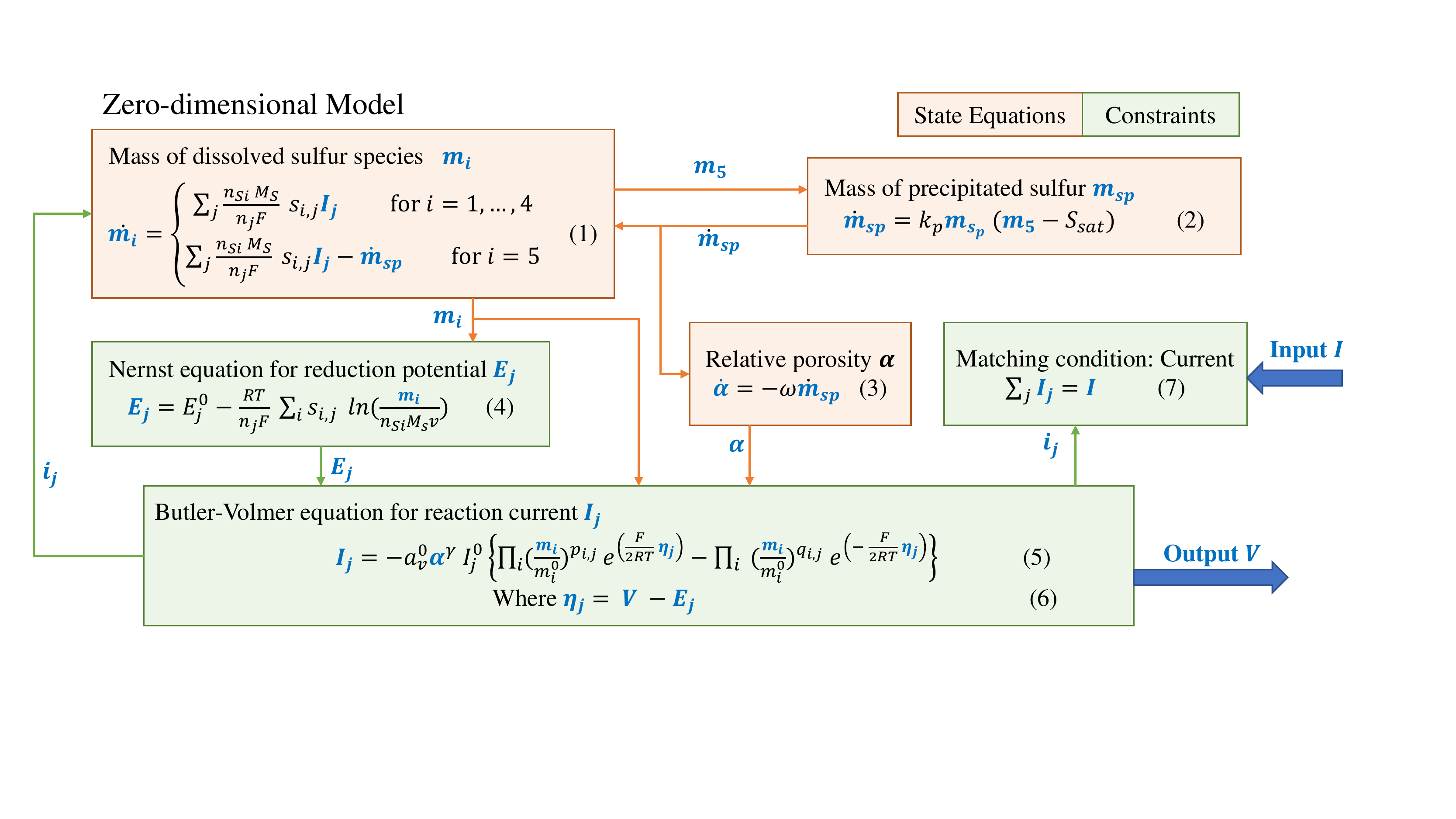}
    \caption{The structure of the zero-dimensional Li-S battery model}
    \label{fig:model_structure}
\end{figure*}

The model adopted in this work captures the physics of key reactions and precipitation phenomena in Li-S batteries. Moreover, by its very nature as a zero-dimensional model, it neglects diffusion/migration dynamics for simplicity. These dynamics lead to the time evolution of the model's state variables, namely: the masses of the various sulfur species and the porosity of the cathode material. The resulting differential algebraic equation (DAE) model (where positive current denotes discharge) is shown in Fig \ref{fig:model_structure}, including the state equations (Eqn. 1-3), algebraic constraints (Eqn. 4-7) and the information flow.  

The input of the system is current $I$, and the output of the system is the voltage measurement across the battery $V$. The state variables include: the masses of the dissolved sulfide species $m_i$ ($i = 1,...,5$), the mass of the precipitated sulfur $m_{S_p}$, and the relative porosity of the cathode material $\alpha$. The rates of mass change for higher-order dissolved sulfide species relate to the current $I_j$ generated by reaction $j$ ($j=1,...,4$). For the last dissolved species $S^{2-}$, one needs to consider its mass generation from the last reduction reaction and its mass loss due to precipitation as in Eqn. 1, where $i$ and $j$ are the indices of the species and reactions. The nucleation and growth phenomenon is described by Eqn. 2. The parameters $k_p$ and $S_{sat}$ are the precipitation rate constant and the saturation mass of $S^{2-}$. The rate of change of the mass of the precipitate is driven by the precipitated mass. This reflects the fact that the existing precipitate serves as a nucleus for further precipitation/growth, as long as the mass in the electrolyte is above a given saturation mass $S_{sat}$. Relative porosity equals the current porosity of the cathode material divided by initial porosity, and has a direct effect on the active reaction area \cite{kumaresan2008mathematical}. Its rate of change is described in Eqn. 3, with a rate constant $\omega$. When the porosity decreases to zero, all the reactants are blocked by the precipitate from the cathode material surface. This can be one of the indicators of the cell's full discharge, another full discharge scenario being the reduction of all polysulfides to $S^{2-}$. The parameters $n_{S_i}$, $n_j$, $R$, $T$, $v$, $M_s$, $F$ and $s_{i,j}$ represents the number of sulfur atoms in species $i$, number of electrons exchanged in reaction $j$, gas constant, temperature, cell volume, molar mass of a sulfur atom, Faraday’s constant, and stoichiometric coefficients of the reactions, respectively. The coefficients $p_{i,j}$ and $q_{i,j}$ represent the positive and negative elements of $s_{i,j}$.

\begin{table}
\captionsetup{font=scriptsize}
	\begin{center}
		\caption{Key Parameter Values in Simulations}
		\label{tab:para_value}
		\begin{tabular}{|C{2.3cm} |C{5.0cm}| C{2cm}| }
			\hline 
			\textbf{Notations} & \textbf{~Values} & \textbf{Units}  \\ 
			\hline 
			$E_j^0$ & 2.4673,~~2.3742,~~2.3420,~~2.0693 & V \\ \hline
			$I_j^0$ & 2.00,~~0.02,~~0.02,~~0.02 & $A/m^2$ \\ \hline
			$m_i^0$ & 3.0377,~1.83E-05,~1.83E-05  & g \\ 
			   & 1.83E-05,~3.26E-06 &  \\ \hline
  			$v$ & 0.0114 & L \\ \hline
			$S_{sat}$ & 0.0001 & g \\ \hline
			$a_{v}^0$ & 1 & m$^2$ \\ \hline
			$\gamma$ & 0.4832  & - \\ \hline
			$\omega$ & 0.6133 & 1/g \\ \hline
			$k_p$ & 22 & 1/(g~s) \\ \hline
			$M_s$ & 32 &  g/mol\\ \hline
			$n_{S_i}$ & 8,~8,~6,~4,~1&  - \\ \hline
			$n_j$ & 1,~1,~1,~1 & - \\ \hline
			$R$ & 8.3145 & J/(K~mol) \\ \hline
			$F$ & 9.649$\times10^4$ & C/mol \\ \hline
			$T$ & 298 & K \\ \hline
			$s_{ij}$ & $\begin{pmatrix}  
			-1/2 & 0 &  0 &  0 \\
			1/2  & -3/2  &  0  &   0\\
			0   & 2   &  -1   &  0\\
			0   & 0  &  3/2   & -1/6\\
			0  & 0  &  0    &   2/3    
			\end{pmatrix}$ & - \\ 
			\hline
		\end{tabular}
	\end{center}
\end{table}
 
The reduction potential of each reaction $E_j$ is given by the Nerst equation (Eqn. 4), assuming that $E_j^0$ is the corresponding reference potential \cite{zhang2015modeling}. The current generated by the corresponding reduction reaction is described by the Butler–Volmer equation (Eqn. 5), where $\eta_j$ is the overpotential, $I_j^0$ is the exchange current density, $\gamma$ is a morphology parameter serving as the power of the relative porosity and $m_i^0$ is the initial mass of species $i$. All these currents sum to the external discharge current $I$.  The parameters shown in Table \ref{tab:para_value} are obtained from~\cite{Xu2020parameterid}.

\subsection{Model Reformulation}

The Li-S battery model developed in Section 2.1 is a differential algebraic equation (DAE) model. State estimators exist for such models~\cite{boutayeb1995observers,becerra2001applying}, but the fundamental theoretical foundations of estimation theory are more established for traditional explicit ordinary differential equation (ODE) models. One method of converting a DAE model to ODEs is to solve for the algebraic variables explicitly. With this in mind, this section analytically resolves the algebraic loop in the above DAE model, thereby reformulating it into an ODE model. This reformulation also eliminates the need for determining consistent initial conditions for the DAE model, thereby also simplifying the estimation problem. From the information flow described in Fig. \ref{fig:model_structure}, one can identify that solving Eqn. 7 analytically is sufficient for the model reformulation. The goal is to solve for the reaction current $I_j$ in Eqn. 7. To do so, we first solve for the output voltage measurement $V$ using the following steps:

Firstly, substitute Eqn. 4 and 6 into Eqn. 5 and rearrange to get Eqn. \ref{ij_rearrange}:
\setcounter{equation}{7}
\begin{equation}
    I_j = -a_{v}^0 \alpha^\gamma \left( \frac{Y}{\Omega_{1j}}   \prod_i  m_i^{p_{i,j} + \frac{1}{2}s_{i,j}}
    -  \frac{Y^{-1}}{\Omega_{2j}} \prod_i m_i^{q_{i,j} - \frac{1}{2}s_{i,j}} \right) \label{ij_rearrange}
\end{equation}
where
\begin{equation}
  Y = e^{\frac{F}{2RT} V}
\end{equation}
\begin{align}
    \Omega_{1j} &=  \frac{1}{I_j^0}  e^{(\frac{F}{2RT} E_j^0)}  \prod_i  m_i^{p_{i,j}} (n_{S_i} M_s v)^{\frac{1}{2} s_{i,j}} \\
    \Omega_{2j} &=  \frac{1}{I_j^0}  e^{(-\frac{F}{2RT} E_j^0)}  \prod_i  m_i^{q_{i,j}} (n_{S_i} M_s v)^{-\frac{1}{2} s_{i,j}} 
\end{align}
Now, substitute Eqn. \ref{ij_rearrange} into Eqn. 7 to get:
\begin{equation}
    \Delta_1 Y - \Delta_2 Y^{-1} + \frac{I}{a_v} = 0  \label{quadratic_fcn}
\end{equation}
where 
\begin{align}
    \Delta_1 &= \sum_j \left(\frac{\prod_i  m_i^{p_{i,j} + \frac{1}{2}s_{i,j}}}{\Omega_{1j}} \right) \\
    \Delta_2 &= \sum_j \left(\frac{\prod_i  m_i^{q_{i,j} - \frac{1}{2}s_{i,j}}}{\Omega_{2j}} \right)
\end{align}
One can solve for $Y$ from the quadratic function in Eqn. \ref{quadratic_fcn}. Since $Y$ is an exponential function, it is always a positive scalar. This makes it possible to finally determine an analytic solution for the voltage measurement, $V$: 
\begin{equation}
   V = \frac{2RT}{F} \text{ln} \left(\frac{-\frac{I}{a_v}+\sqrt{\frac{I^2}{a_v^2}+4\Delta_1 \Delta_2}}{2\Delta_1} \right)  
\end{equation}
In this way, the algebraic loop is broken, and the resulting model can be expressed in the following standard explicit state-space form:
\begin{align}
  \dot{X} &= f(X,I) \\
   V &= h(X, I)  \label{outputV}
\end{align}
where the state variables form a 7-by-1 vector:
\begin{equation}
  X = [m_1,...,m_5, m_{S_p}, \alpha]^T
\end{equation}

The above model serves as a foundation for the observability analysis and estimation study presented in the remainder of this article. One important note is that the precipitated mass $m_{S_p}$ does not directly affect the output battery voltage $V$. Instead, the impact of precipitation on this output voltage takes place indirectly, through the dynamics of other species masses. This causes the observability of the precipitated mass to be fairly weak, especially in the low plateau region, as shown in the following sections. 

\section{Observability Analysis}
In control theory, observability is an indicator of whether or not the internal state of a dynamic system can be estimated from input/output measurements. For linear systems, the conditions for observability are uniquely/equivalently defined: a linear time invariant system is observable, if and only if the observability matrix is full rank or equivalently the observability gramian is non-singular. In contrast, there are multiple different tests for observability in nonlinear systems~\cite{hermann1977nonlinear}. This section analyzes the so-called ``weak observability" of the Li-S ODE model, defined as one's ability to estimate the model's state within a local neighborhood of its true value. This analysis furnishes a Cram\'er-Rao bound on the best-achievable state estimation accuracy for any give charge/discharge profile and time period, assuming a known measurement noise distribution. The first goal of this analysis is to confirm that the model is locally observable, thereby justifying the subsequent development of an estimator. The second goal is to compare observability for different state variables during different phases of battery operation (e.g., high plateau, low plateau, etc.).   

One method to solve the problem of estimating the current state of a dynamic system can be done in two steps: (i) estimating the system's initial state and (ii) propagating the initial state though the system model to obtain the current state. The empirical observability gramian adopts this idea~\cite{powel2015empirical}. The observability gramian is calculated by perturbing the initial state $X_0$ along each state variable by a positive small value $\varepsilon$, and then simulating the system for a time period $\tau$ to get the resulting differences in output measurements. This furnishes the following equation for the observability gramian: 
\begin{equation}
    W_o^\varepsilon (\tau, X_0, I) = \frac{1}{\varepsilon^2} \int_0^ \tau \Phi^\varepsilon(t,X_0,I)^T \Phi^\varepsilon(t,X_0,I) dt
    \label{obsv_gramian}
\end{equation}
where
\begin{align}
    \Phi^\varepsilon(t,X_0,I) &= [V^{+1} - V^{1},  ~~ ...~~, V^{+n} - V^{n}] \\
   V^{+i} &= h(X(t, X_0 +\varepsilon e_i),I) \\
   V^{i} &= h(X(t, X_0),I) \text{~~~~for $i=1,...,n$}
\end{align}

The vectors $e_i$ represent the elements of the standard basis in the state variable domain $\mathbb{R}^n$ (e.g., $e_1 = [1~0~0~0~0~0~0]^T$), and $n$ equals 7. If the observability gramian is non-singular, the nonlinear system is locally observable, which means that one can estimate the initial values of its state variables from input/output measurements. The best theoretically achievable accuracy with which such estimation can be performed is governed by the associated Fisher information matrix~\cite{gorman1990lower}. This matrix is formally derived in~\cite{norton2009introduction}. Moreover, assuming that the voltage output measurement noise process is white and Gaussian, this Fisher information matrix can be simplified to the following expression: 
\begin{equation}
    F_M =  \frac{1}{\sigma_v^2} \sum^N_{k=1} \left[ \left(\frac{\partial \mathbf{V}}{\partial X_0} \right) \left(\frac{\partial \mathbf{V}}{\partial X_0} \right)^T \right] 
\end{equation}
In the above equation, $N$ is the number of samples within the time period $\tau$, (boldface) $\mathbf{V}$ is a 1-by-N row vector of voltage measurements at every time step, and  the sampling time step is $\delta t = \frac{\tau}{N}$. When the sampling time step is small enough, the above equation for the Fisher information matrix can be approximated using the following integral \cite{sharma2014fisher}:
\begin{align}
 \begin{split}
F_M &= \frac{1}{\sigma_v^2~\delta t}  \int_0^\tau \left(\frac{\partial V}{\partial X_0} \right) \left(\frac{\partial V}{\partial X_0}  \right)^T dt \ \\
 &= \frac{1}{\sigma_v^2 ~\delta t} W_o^\varepsilon (\tau, X_0, I)
 \end{split}
 \label{FisherMatrix}
\end{align}

When the Fisher matrix is invertible (i.e. the system is locally observable), then the inverse of the matrix provides a Cram\'er-Rao lower bound (CRLB) that represents the best achievable estimation error covariance for the estimated initial state vector $X_0$. 
\begin{equation}
cov(X_0) \geq F_M^{-1} = \sigma_v^2~\delta t~W_o^\varepsilon (\tau, X_0, I) ^{-1}
\label{CramerR}
\end{equation}

Due to the nonlinearity and complexity of our system model, we evaluate the observability gramian and the Cram\'er-Rao lower bound numerically along the system's discharge trajectory, given a constant discharge C-rate of 0.3C. The parameters used in this observability analysis are the perturbation $\varepsilon = 10^{-6}$, the sampling time step $\delta t = 0.1$ second, the time period $\tau = 60$ seconds and the standard deviation of the output measurement $\sigma_v = 5\times 10^{-3}$ V. 

\begin{figure}
    \centering
    \includegraphics[trim={0.5cm 0cm 0.5cm 0cm},clip, width= 0.7\textwidth]{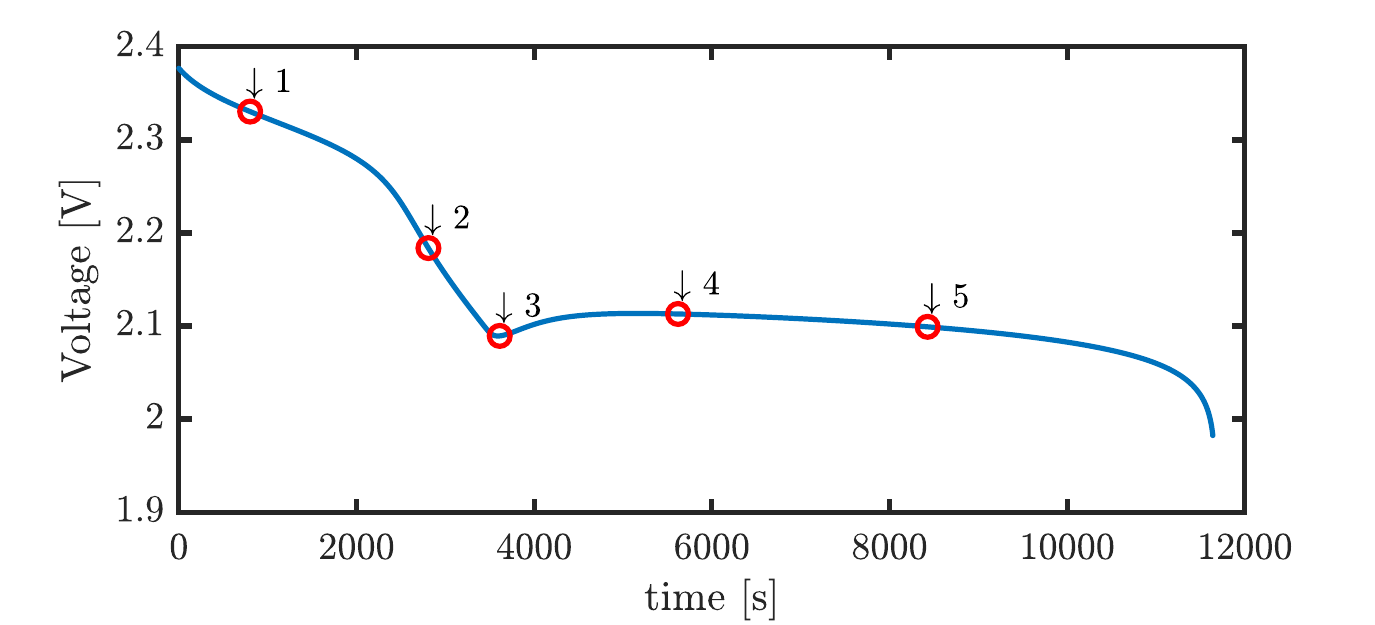}
    \caption{Observability check points along the voltage profile}
    \label{fig:CheckPoint_Voltage}
\end{figure}

We choose five representative points (two points at each plateau and the middle dip point) to obtain the empirical observability gramian and CRLB along the voltage measurement profile as shown in Fig \ref{fig:CheckPoint_Voltage}. At all these 5 points, the empirical observability gramian is invertible, leading to the conclusion that the system is indeed locally (or ``weakly") observable. This means that if the estimated initial states are in the neighborhood of the true values, one can construct an unbiased observer for the system (i.e., an observer that converges to the true values in an average statistical sense). The best statistical accuracy (i.e., statistical spread of initial state estimates) achievable by this observer is given by the CRLB. Specifically, the square roots of the diagonal elements of the CRLB matrix represent the standard deviations (std) of the best achievable estimation error for each initial state variable. These standard deviations are listed in Table \ref{tab:std_checkpoint_ori}. 

\begin{table}
\captionsetup{font=scriptsize}
	\begin{center}
		\caption{Standard deviation of the best achievable estimation error for the full-order model}
		\label{tab:std_checkpoint_ori}
		\begin{tabular}{|C{1.5cm}|C{1.2cm}|C{1.2cm}|C{1.2cm}|C{1.2cm}|C{1.2cm}| }
			\hline 
			\textbf{std [g]} & \textbf{Nr. 1} & \textbf{Nr. 2}& \textbf{Nr. 3}& \textbf{Nr. 4}& \textbf{Nr. 5} \\\hline
    \bm{$m_1$}  & 0.028 & 	0.014 & 	0.006 & 	0.039 & 	0.029  \\ \hline 
    \bm{$m_2$}  & 0.022 & 	0.018 & 	0.006 & 	0.031 & 	0.022  \\\hline 
    \bm{$m_3$}  & 0.014 & 	0.017 & 	0.009 & 	0.094 & 	0.088  \\\hline 
    \bm{$m_4$}  & 0.017 & 	0.016 & 	0.006 & 	0.092 & 	0.096  \\\hline 
    \bm{$m_5$} & 0.002 & 	0.003 & 	0.005 & 	0.063 & 	0.072  \\\hline 
    \bm{$m_{S_p}$}  & 0.076 & 	0.071 & 	0.046 & 	0.282 & 	0.367 \\\hline 
    \bm{$\alpha$}  & 0.033 & 	0.024 & 	0.009 & 	0.098 & 	0.106  \\\hline 
		\end{tabular}
	\end{center}
\end{table}

Two main observations are visible from Table \ref{tab:std_checkpoint_ori}. First, the best local observability shows up at the dip point between the high and low plateau. Estimation errors are slightly worse in the high plateau compared to this dip point, and the worst estimation errors occur in the low plateau. This observation matches the intuition that the estimation problem is more difficult during the low plateau due to the associated flatness of the measured voltage profile. The smaller sensitivity of the output voltage with respect to underlying state variables in the lower plateau regions increases the difficulty of the state estimation problem, and therefore worsens estimation accuracy. Second, the largest standard deviation of the state estimation error corresponds to the mass of precipitated sulfur $m_{S_p}$. This is especially problematic in the low plateau region, where this error can be as large as 0.367 g (compared to the total active sulfur mass of 3.0377 g). This is due to the fact that the output voltage is not directly governed by the precipitated mass $m_{S_p}$, i.e. the variable $m_{S_p}$ does not show up in the output equation, Eqn. ~\ref{outputV}. In practice, when performing state estimation during the low plateau, if the initial estimated $m_{S_p}$ is not close to the true value, this error will persist. 

\subsection{Improving Observability through Model Reduction}

When a dynamic system suffers from poor observability, one potential remedy is to estimate a subset of the system's state variables. This subset can be selected to ensure that it has attractive observability properties. Estimates of the remaining state variables must then be computed through other means, such as the exploitation of prior information. The principle of mass conservation provides an opportunity for improving the observability of this article's Li-S battery model. The basic idea is to assume that the total active sulfur mass is accurately known \textit{a priori}. Prior knowledge of this mass can be obtained from a number of different sources, one of which is state estimation during the high plateau. Armed with this prior knowledge, one can construct a reduced-order model where the principle of mass conservation is used to eliminate two state variables, namely: the precipitated sulfur mass and relative porosity. One can then construct an estimation algorithm for the remaining 5 state variables using the resulting reduced-order model.  

Consider the mass conservation of all the sulfur species. The rate of change of the total sulfur mass is zero. One can therefore eliminate the state variable $m_{S_p}$ as follows:
\begin{equation}
    m_{S_p} = M_{tot} - \sum_{i=0}^5 m_i \label{precip_sulfur}
\end{equation}
where $M_{tot}$ represents the total sulfur mass, known \textit{a priori}. Moreover, knowing that the initial relative porosity for a fully charged cell equals 1 by definition, and assuming the corresponding precipitated sulfur mass to be approximately 0, one can further reduce the state variable $\alpha$ from the model as follows:
\begin{equation}
    \alpha = 1- \omega (M_{tot} - \sum_{i=0}^5 m_i)
\end{equation}

The resulting reduced-order ODE model contains 5 state variables. The authors perform the same observability analysis to obtain the standard deviations of the corresponding best achievable estimation errors. The results of this analysis are listed in Table \ref{tab:std_checkpoint_reduce}. The observability of the masses of the dissolved species in the 5th-order model is on the same order of magnitude, and generally better, compared to the 7th-order model. More importantly, relative porosity and precipitated sulfur mass are no longer being directly estimated in this reduced-order model, the result being that their estimation accuracy is now a strong function of the fidelity of the prior estimate of total sulfur mass. Accurate prior knowledge of this total mass can therefore benefit the estimation process, as demonstrated in the following sections of the article. The remainder of this article develops and simulated an unscented Kalman filter for estimating the state variables of both of the above 7th-order full model and 5th-order reduced model. 

\begin{table}
\captionsetup{font=scriptsize}
	\begin{center}
		\caption{Standard deviation of the best achievable estimation error for the reduced-order model}
		\label{tab:std_checkpoint_reduce}
		\begin{tabular}{|C{1.5cm}|C{1.2cm}|C{1.2cm}|C{1.2cm}|C{1.2cm}|C{1.2cm}| }
			\hline 
			\textbf{std [g]} & \textbf{Nr. 1} & \textbf{Nr. 2}& \textbf{Nr. 3}& \textbf{Nr. 4}& \textbf{Nr. 5} \\\hline
    \bm{$m_1$}   & 0.007 & 	0.009 & 	0.005 & 	0.027 & 	0.030  \\ \hline 
    \bm{$m_2$}   & 0.021 & 	0.015 & 	0.006 & 	0.024 & 	0.020  \\\hline 
    \bm{$m_3$}   & 0.014 & 	0.013 & 	0.006 & 	0.059 & 	0.088  \\\hline 
    \bm{$m_4$}   & 0.015 & 	0.009 & 	0.006 & 	0.052 & 	0.071  \\\hline 
    \bm{$m_5$}   & 0.002 & 	0.002 & 	0.003 & 	0.062 & 	0.056  \\\hline 
		\end{tabular}
	\end{center}
\end{table}

\section{State Estimation using UKF}

Kalman filtering is a popular and well-established approach for state/parameter estimation. The traditional Kalman filter provides state estimates for linear dynamic systems, but nonlinear extensions of this filter exist, including both the extended and unscented Kalman filters. Compared to the extended Kalman filter (EKF), the use of sigma points for propagating estimation covariance in unscented Kalman filters generally improves estimation accuracy~\cite{wan2000unscented,wan2001unscented,sarkka2007unscented}. This section applies unscented Kalman filtering to the full and reduced-order Li-S ODE models, and evaluates the performance of the resulting filters in simulation. 

\subsection{The UKF Algorithm}
The state and output equations of the nonlinear Li-S battery model can be discretized in time and augmented with both process and measurement noise signals to give: 

\begin{align}
    X_k &= f(X_{k-1}, I_k) + v_k \label{mass_ode_discrete}  \\
    V_k &= h( X_k, I_k) + w_k
\end{align}
In the above discrete-time model, $v_k$ and $w_k$ are the process and measurement noise, respectively, which are both uncorrelated zero-mean Gaussian white sequences with known variances. Then, the UKF algorithm is shown as follows.\\
\textbf{(1) Initialization:}

Define an initial state estimate $\hat{X_0}$ and covariance matrix $P_0$. In this work, the initial covariance matrix $P_0$ is set as $ diag ( P_0 )=  0.1 \hat{X}_0$. \\
\textbf{(2) Generate sigma points at time step k-1:}
\begin{align}
     \chi_{k-1}^0 &= \hat{X}_{k-1} \\
     \chi_{k-1}^i &= \hat{X}_{k-1} + \sqrt{(N+\lambda)P_{k-1}} ~~~i=1,...,N \\
     \chi_{k-1}^i &= \hat{X}_{k-1} - \sqrt{(N+\lambda)P_{k-1}} ~~~i=N+1,...,2N
\end{align}
where $\lambda$ is a scaling factor    
\begin{equation}
   \lambda = \beta^2(N+\kappa)-N
\end{equation}
Here $\beta \in [0,1]$ and $\kappa \in [0,\infty]$ are two tuning parameters that determine the spread of the sigma points. In this study, we choose 
$\beta =0.01$ and $\kappa= 1$.\\
\textbf{(3) Time update:}

The sigma points are propagated through the nonlinear discrete state equation to obtain the estimated state matrix $\chi_{k}^{i-}$
\begin{equation}
    \chi_{k}^{i-} =f( \chi_{k-1}^i, I_k) ~~~i=0,...,2N
\end{equation}

\noindent The \textit{a priori} state estimate is then given by: 
\begin{align}
    \hat{X}_{k}^-  &=\sum_{i=0}^{2N} W_i^m \chi_{k}^{i-} \\
     W_0^m &=\frac{\lambda}{\lambda+N} \\
     W_i^m &=\frac{\lambda}{2\lambda+2N} ~~~i=1,...,2N
\end{align}

\noindent Moreover, the \textit{a priori} error covariance is calculated as
\begin{align}
    P_{k}^-  &=\sum_{i=0}^{2N} W_i^c [\chi_{k}^{i-} -\hat{X}_{k}^- ][\chi_{k}^{i-} -\hat{X}_{k}^- ]^T + Q_k\\
     W_0^c &=\frac{\lambda}{\lambda+N} + (1-\beta^2-\mu) \\
     W_i^c &=\frac{\lambda}{2\lambda+2N} ~~~i=1,...,2N
\end{align}

In this work, a constrained version of the UKF which considers allowable limits on species mass values is implemented. The sizing of the process noise covariance matrix is used for adjusting the spread of the UKF sigma points. It does not necessarily imply or preclude the existence of substantial process noise in the physical system. A bigger covariance matrix leads to a broader spread of the sigma points. If $Q_k$ is too big, it is possible to generate infeasible sigma points with negative species mass values. Hence, the authors enforce a cap on the maximum values of the diagonal elements of $Q_k$ as below:
\begin{equation}
   diag ( Q_k )=  min [0.005, ~~0.005 \hat{X}_{k-1}]
\end{equation}
The estimated measurement matrix  $Z_k$, is calculated by transforming the sigma points using the nonlinear discrete output equation
\begin{equation}
    Z_{k}^{i} =h(\chi_{k-1}^i, I_k) ~~~i=0,...,2N
\end{equation}

\noindent Moreover, the estimated measurement $\hat{V}_k$ is given by: 

\begin{equation}
    \hat{V}_{k}  =\sum_{i=0}^{2N} W_i^m Z_{k}^{i} \\
\end{equation}
\textbf{(4) Measurement update:}

The Kalman gain $K_k$ is calculated from the measurement covariance $P_{z}$ and the cross-correlation covariance $P_{xz}$
\begin{align}
    P_{z}  &=\sum_{i=0}^{2N} W_i^c [Z_{k}^{i} -\hat{V}_{k} ][Z_{k}^{i} -\hat{V}_{k}]^T + R_k\\
    P_{xz}  &=\sum_{i=0}^{2N} W_i^c [\chi_{k}^{i-} -\hat{X}_{k}^- ][Z_{k}^{i} -\hat{V}_{k}]^T \\
    K_k  &= P_{xz} P_{z}^{-1}
\end{align}
where $R_k = \sigma_v^2$, with the standard deviation of the output measurement $\sigma_v = 5\times 10^{-3}$ V defined as in Section III. 

The state estimate is updated by the measurement
\begin{equation}
    \hat{X}_{k} = \hat{X}_{k}^- + K_k (V_k - \hat{V}_k)
\end{equation}

Finally, the error covariance matrix is updated through
\begin{equation}
    P_{k}  = P_{k}^-   - K_k P_{z}  K_k^T
\end{equation}

\begin{figure*}
    \centering
    \includegraphics[trim={8cm 3.5cm 8cm 3.2cm},clip, width= 0.97\textwidth]{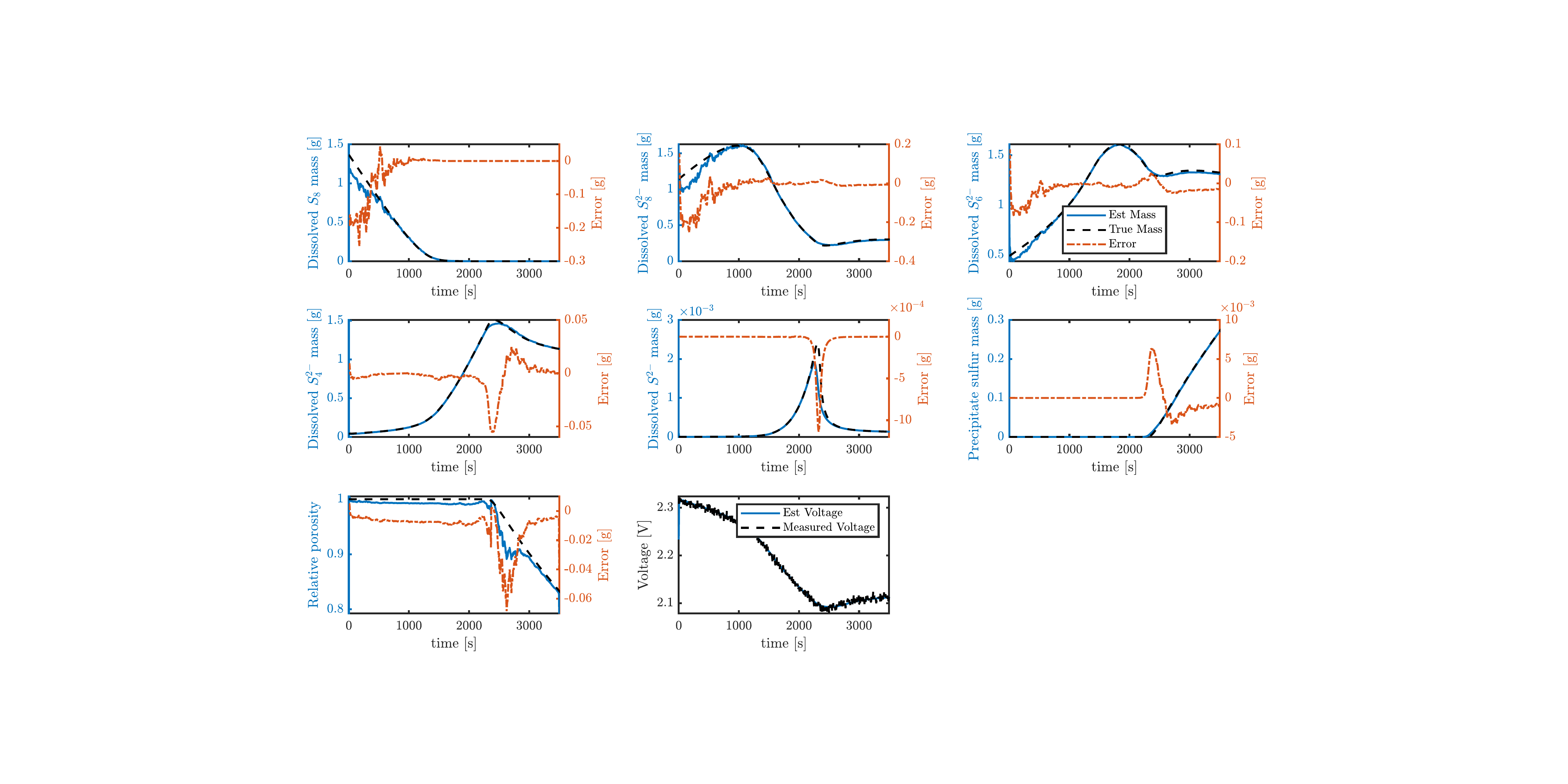}
    \caption{High plateau state estimation with a constant discharge current for the full-order model}
    \label{fig:highPlateau_est_const_cur_full}
\end{figure*}
\begin{figure*}
    \centering
    \includegraphics[trim={8cm 3.5cm 8cm 3.2cm},clip, width= 0.97\textwidth]{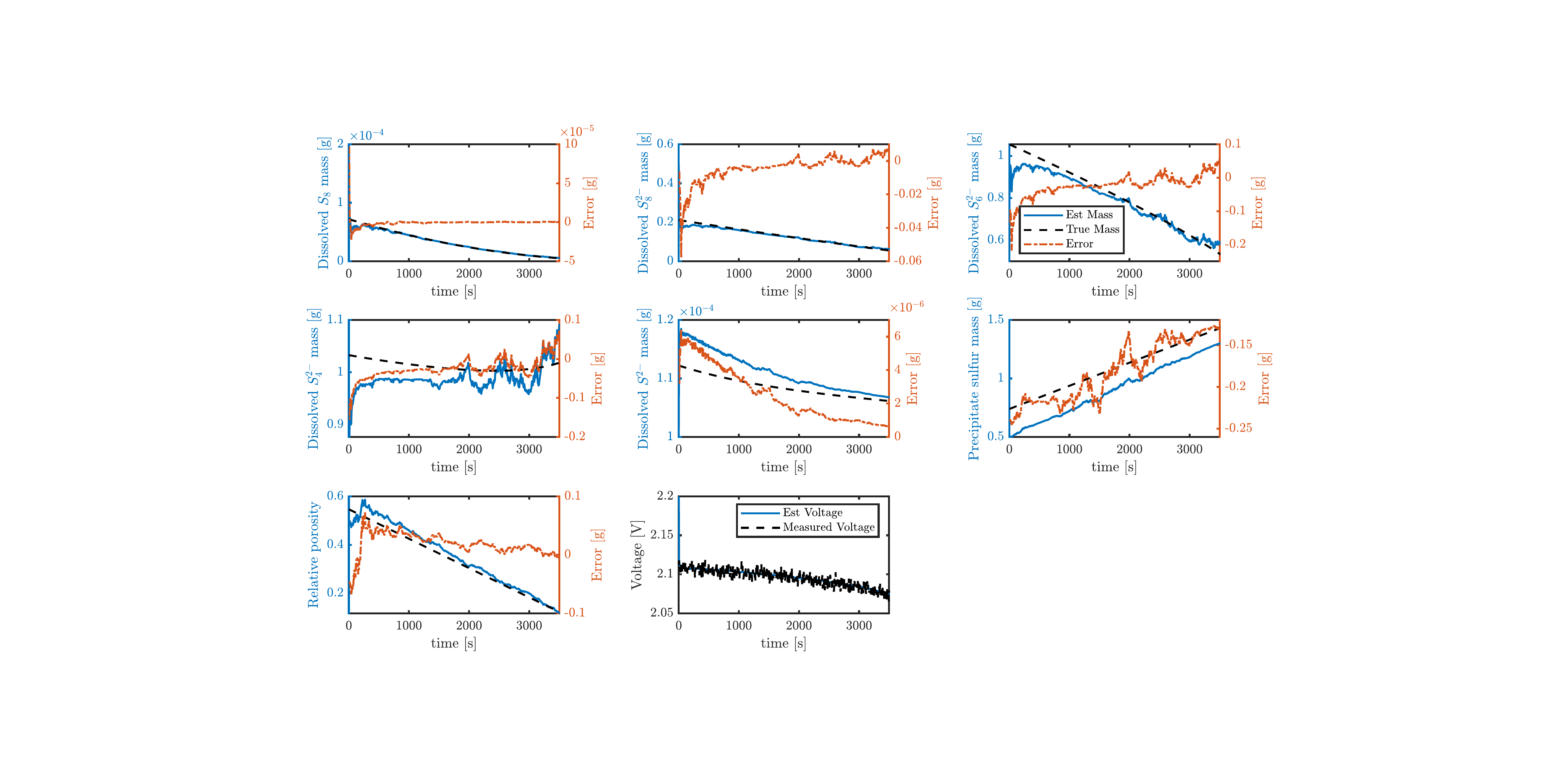}
    \caption{Low plateau state estimation with a constant discharge current for the full-order model}
    \label{fig:lowPlateau_est_const_cur_full}
\end{figure*}

\subsection{Simulation Results and Discussions}
The UKF algorithm is tested both in the high and low plateau regions in simulation, for both the full- and reduced-order models. Although state estimation is performed at different check points, consistent results are observed for points located in the same plateau region. The authors therefore choose one initial state for each plateau as representatives for demonstration and discussion. For the full-order ODE model, a constant current discharge scenario is tested with the discharge current set to 1 Ampere (corresponding to a C-rate of 0.3C). This simulated scenario is repeated for the reduced-order model. The article also simulates the reduced-order model for a sinusoidal current input, as discussed below.

\subsubsection{State Estimation for Full-order ODE Model}

Fig. \ref{fig:highPlateau_est_const_cur_full} and \ref{fig:lowPlateau_est_const_cur_full} show the simulation results for the full-order ODE model with a constant current input of 1 Ampere. In the high plateau region, the masses of the three main species $S_8$, $S_8^{2-}$ and $S_6^{2-}$ demonstrate good convergence. The masses of both the sulfur precipitate and dissolved $S^{2-}$ are small in this region, and so are the UKF estimates of these masses. Broadly speaking, therefore, the performance of the full-state UKF estimator in the high plateau region is reasonable. This is consistent with the article's earlier analysis showing good state observability in this region. 

An obvious disadvantage shows up in the estimation of precipitated sulfur mass: this precipitated mass is estimated with poor accuracy, especially in the low plateau region. Errors in estimating this precipitated mass do not converge to negligible values even after an hour of battery discharge, as shown in Fig. \ref{fig:lowPlateau_est_const_cur_full}. This can be explained by the observability analysis earlier in this article. The output voltage measurement only relates to the dissolved sulfur species, as they affect the associated reduction potentials through the Nernst equation. When performing state estimation in the low plateau region, if the initial estimated $m_{S_p}$ is not close to the true value, the error will persist without substantial improvement. This is a fundamental issue, rather than a problem specific to the filtering algorithm used in this article, as evident from the earlier observability analysis.

\subsubsection{State Estimation for Reduced-order ODE Model}

\begin{figure*}
    \centering
    \includegraphics[trim={8cm 2cm 8cm 2cm},clip, width= 1\textwidth]{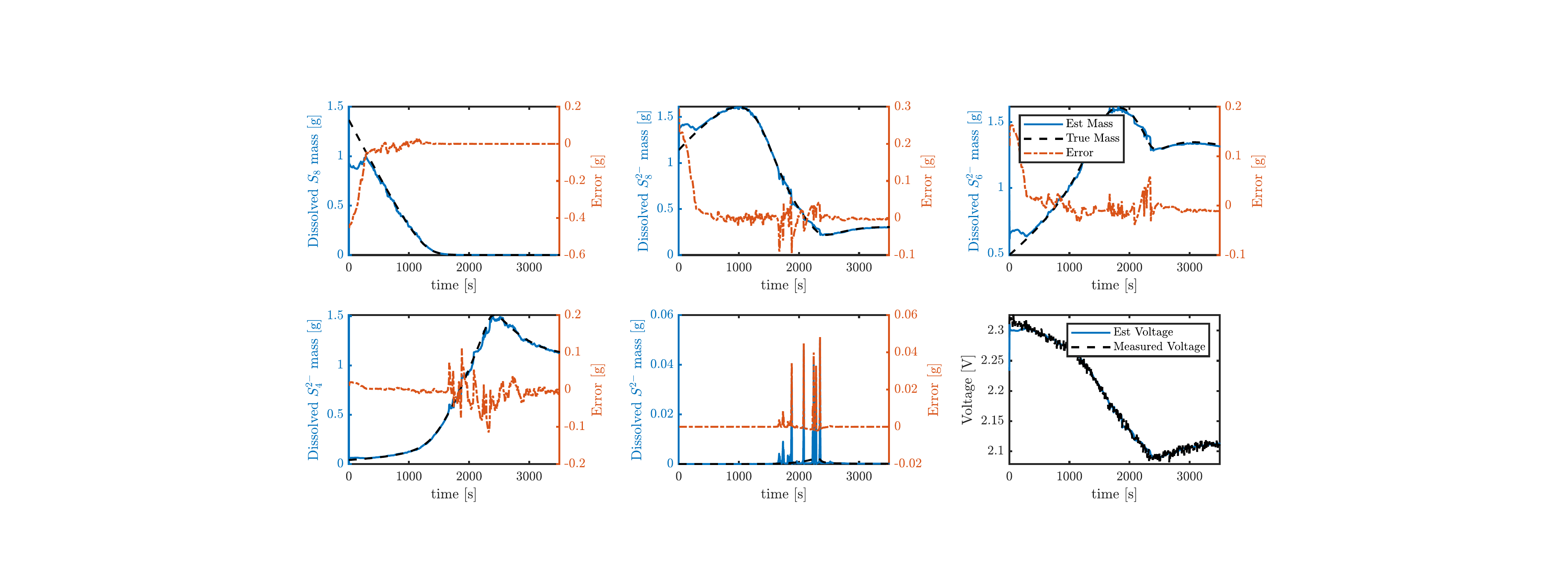}
    \caption{High plateau state estimation with a constant discharge current for the reduced-order model}
    \label{fig:highPlateau_est_const_cur}
\end{figure*}
\begin{figure*}
    \centering
    \includegraphics[trim={8cm 2cm 8cm 2cm},clip, width= 1\textwidth]{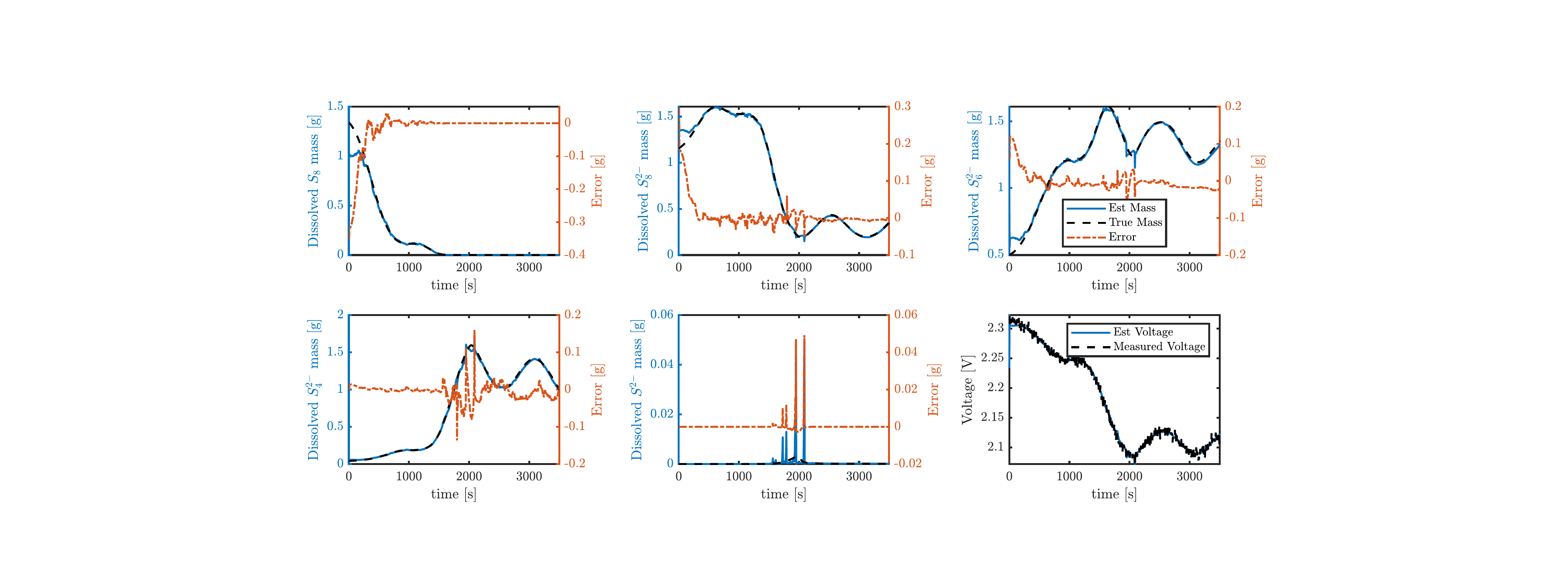}
    \caption{High plateau state estimation with a constant plus sinusoidal discharge current for the reduced-order model}
    \label{fig:highPlateau_est_sin_cur}
\end{figure*}

Two simulation scenarios are examined for unscented Kalman filtering using the reduced-order battery model, namely: a 1-Ampere constant-current discharge scenario and a sinusoidal input current scenario with $I = 1+ sin(0.005t)$ Ampere. The simulation results for high plateau estimation with both of these current profiles are shown in Fig. \ref{fig:highPlateau_est_const_cur} and \ref{fig:highPlateau_est_sin_cur}, respectively. One can observe the state estimates converge to the true value within about 5 minutes. The slow-changing current input does not lead to significant change in the convergence rate.

\begin{figure*}
    \centering
    \includegraphics[trim={8cm 2cm 8cm 2cm},clip, width= 1\textwidth]{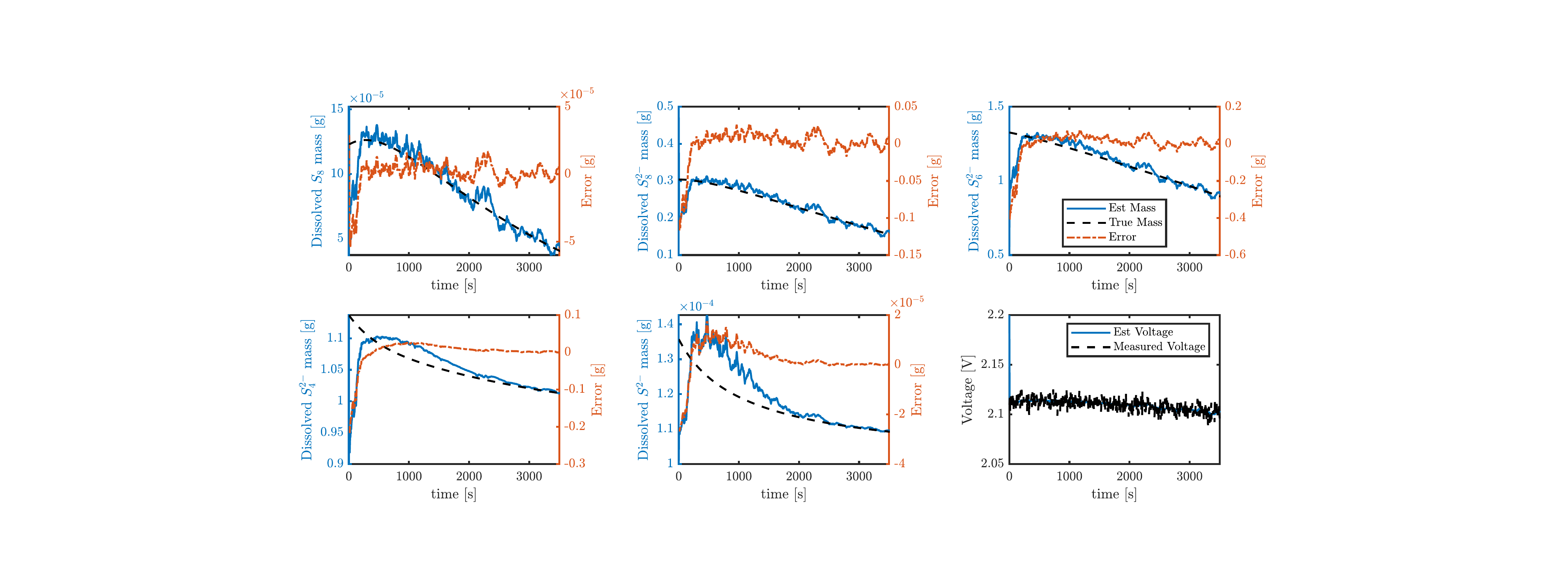}
    \caption{Low plateau state estimation with a constant discharge current for the reduced-order model}
    \label{fig:lowPlateau_est_const_cur}
\end{figure*}

\begin{figure*}
    \centering
    \includegraphics[trim={8cm 2cm 8cm 2cm},clip, width= 1\textwidth]{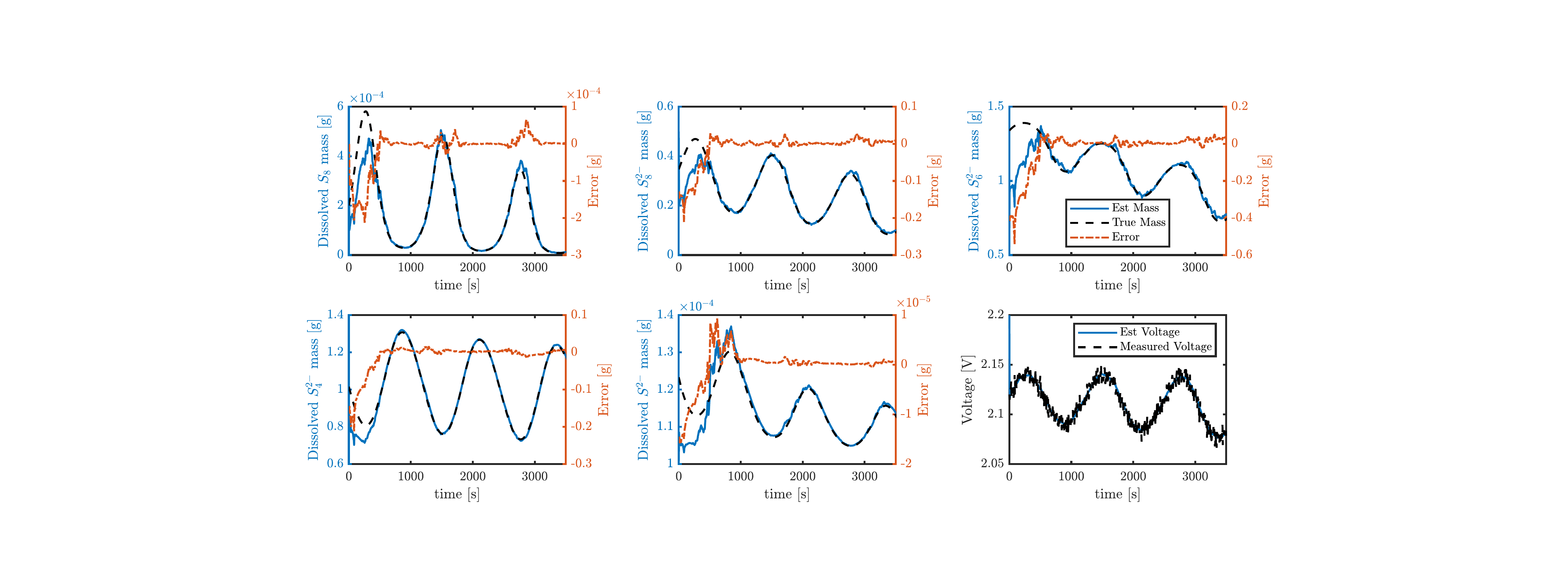}
    \caption{Low plateau state estimation with a constant plus  sinusoidal discharge current for the reduced-order model}
    \label{fig:lowPlateau_est_sin_cur}
\end{figure*}

Fig. \ref{fig:lowPlateau_est_const_cur} and \ref{fig:lowPlateau_est_sin_cur} present the estimation results for the low plateau. The convergence of the estimator takes more time in the low plateau region. This is because the low plateau manifests less observability, and the estimator requires a longer time duration to gather information. This poor observability can be explained in terms of the reduced sensitivity of the output voltage to the values of the underlying states (i.e. the ``flatness" of the output voltage curve in the low plateau region). Significant differences between the initial state estimates and the corresponding true values take longer to diminish because of this reduced sensitivity. This slow convergence occurs even when the estimated voltage quickly converges to the measured voltage: a sign of poor observability. Intuitively, because of the poor sensitivity of the output voltage with respect to the underlying state estimates, one cannot rely on the voltage measurement error to achieve fast convergence of the state estimates.

Regardless of the above caveats, state estimation accuracy using the reduced-order model is generally attractive. Better observability and estimation accuracy is achievable in the high plateau region, compared to the low plateau region, for both the full and reduced-order estimators. Therefore, it is recommended to launch estimation in the high plateau region if possible, particularly if the total active sulfur mass is unknown. Then one can use the information gathered in the high plateau region to estimate this total active mass and construct a reduced-order state estimator for subsequent time periods. 

\section{Conclusions}
This article shows that it is generally possible to construct an algorithm for estimating the masses of various species in Li-S batteries. This is important because it provides a more detailed picture of the internal state of the battery compared to a more traditional ``lumped" SOC estimate. Observability analysis reveals two fundamental challenges associated with such state estimation. The first challenge is that in the low plateau region, the poor sensitivity of output voltage to species masses results in poor observability and slow estimator convergence. The second challenge is that the mass of precipitated sulfur suffers from particularly poor observability compared to other species masses. These two problems compound, in the sense that estimating precipitate mass is especially challenging in the low plateau region. Under the assumption that the total active sulfur mass is known \textit{a priori}, model reduction makes it possible to circumvent these challenges and obtain accurate state estimates. This raises the possibility of perhaps using switching schemes where full state estimation is performed in the high plateau region, and reduced-order state estimation is performed in the low plateau region. The use of unscented Kalman filtering for estimating species masses in an Li-S battery furnishes simulation results that are consistent with the fundamental discoveries of this article's observability analysis. 

\section*{Acknowledgment}
This work is funded by National Science Foundation Grant 1351146. The authors gratefully acknowledge this support. Any opinions, findings, and conclusions or recommendations expressed in this material are those of the authors and do not necessarily reflect the views of the National Science Foundation.


\bibliography{LiS_state_estimation}

\end{document}